\documentclass[12pt,preprint]{aastex}

\long\def\symbolfootnote[#1]#2{\begingroup%
\def\thefootnote{\fnsymbol{footnote}}\footnote[#1]{#2}\endgroup}

\begin{document}

\title{Stellar Structure of Dark Stars: a first phase of Stellar Evolution
resulting from  Dark Matter Annihilation}
\author{
Katherine Freese\altaffilmark{1},
Peter Bodenheimer\altaffilmark{2},
Douglas Spolyar\altaffilmark{3},
and
Paolo Gondolo\altaffilmark{4}}

\email{ktfreese@umich.edu, peter@ucolick.org, dspolyar@physics.ucsc.edu, paolo@physics.utah.edu}

\altaffiltext{1}{Michigan Center for Theoretical Physics, Physics Dept.,
Univ. of Michigan, Ann Arbor, MI 48109}
\altaffiltext{2}{Dept. of Astronomy and Astrophysics, University of California,
Santa Cruz, CA 95064}
\altaffiltext{3}{Physics Dept., University of California, Santa Cruz, CA 95064}
\altaffiltext{4}{Physics Dept., University of Utah, Salt Lake City, UT 84112}

\begin{abstract}
%\noindent
Dark Stars are the very first phase of stellar evolution in the
history of the universe: the first stars to form (typically at redshifts $z \sim
10-50$) are powered by heating from dark matter (DM) annihilation
instead of fusion (if the DM is made of particles which are their own
antiparticles).  We find equilibrium polytropic configurations for
these stars; we start from the time DM heating becomes important ($M
\sim 1-10~M_\odot$) and build up the star via accretion up to 1000~
M$_\odot$.  The dark stars, with an assumed particle mass of 100 GeV,
are found to have luminosities of a few times $10^6$ L$_\odot$,
surface temperatures of 4000--10,000 K, radii $\sim 10^{14}$ cm,
lifetimes of at least $ 0.5$ Myr, and are predicted to show lines of
atomic and molecular hydrogen.  Dark stars look quite different from
standard metal-free stars without DM heating: they are far more massive
(e.g. $\sim 800 M_\odot$ for 100 GeV WIMPs), cooler, and larger,
and can be distinguished in future observations, possibly even by JWST or TMT.

\end{abstract}

\section{Introduction}
The first stars in the Universe mark the end of the cosmic dark ages,
reionize the Universe, and provide the enriched gas required for later
stellar generations.  They may also be important as precursors to
black holes that coalesce and power bright early quasars.  The first
stars are thought to form inside dark matter (DM) halos of mass $ 10^5
M_\odot$--$ 10^6 M_\odot$  at redshifts $z \sim 10-50$ (Yoshida et
al. 2003).  These halos consist of 85\% DM and 15\% baryons in the
form of metal-free gas made of H and He.  Theoretical calculations
indicate that the baryonic matter cools and collapses via H$_2$
cooling (Peebles \& Dicke 1968, Matsuda et al. 1971, Hollenbach \&
McKee 1979) into a single small protostar (Omukai \& Nishi 1998) at
the center of the halo (for reviews see Ripamonti \& Abel 2005;
Barkana \& Loeb 2001; Bromm \& Larson 2004).

Previously, Spolyar et al. (2008; hereafter, Paper I) first considered the
effect of DM  particles on the first stars during their
formation.  Any DM particle which is capable of annihilating with
itself in such a way as to give the correct relic abundance today will
also annihilate wherever the DM density is high.
The first protostars and stars are particularly good sites for annihilation
because they form at high redshifts (density scales as $(1+z)^3$) and
in the high density centers of DM haloes.  Paper I  found that
DM annihilation provides a powerful heat source in the first stars, a
source so intense that its heating overwhelms all cooling mechanisms.
Paper I suggested that the
very first stellar objects might be {\it Dark Stars} (DS), a new phase of
stellar evolution in which the DM -- while only a negligible fraction
of the star's mass -- provides the key power source for the star through
DM heating. Note that the term 'Dark' refers to the power source, not the
luminosity.  In this paper, we continue the work originally suggested 
in Paper I by studying the DS structure.      
%We find their  mass, luminosity, temperature, and radius. 

The canonical example of particle DM is Weakly Interacting Massive
Particles (WIMPs), which automatically provide the right amount of DM,
i.e.\ $\sim$ 24\% of the current energy density of the Universe.  In
many theories WIMPs are their own antiparticles and annihilate with
themselves in the early universe, leaving behind this relic
density. In particular, the neutralino, the supersymmetric partner of
the W, Z, and Higgs bosons, is a strong candidate (reviewed by Jungman
et al. 1996).  As our canonical values, we use the standard 
$\langle \sigma v \rangle = 3 \times 10^{-26}\,{\rm cm^3/s}$ for the
annihilation cross section and $m_\chi = 100\,{\rm GeV}$ for the 
particle mass.  A companion paper will generalize to other masses and
cross sections. The analysis in this paper could apply equally well to
other DM candidates.

WIMP annihilation produces energy at a rate per unit volume 
\begin{equation}
\hat Q_{DM} = \langle \sigma v \rangle \rho_\chi^2/m_\chi ,
\label{eq:Q}
\end{equation}
where $\rho_\chi$ is the energy density of the WIMPs.
In the early stages of Pop III star formation, when the gas density is
low ($n \lesssim 10^4 {\rm cm}^{-3}$), most of the annihilation products simply
escape from the protostar without heating it
(Ripamonti et al. 2007). However, 
a crucial transition takes place (Paper I)  when the gas density
of the collapsing protostar exceeds a critical value at which point
most of the annihilation energy is trapped in the star.  For a 100
GeV particle,  at hydrogen density  $\sim 10^{13}\, {\rm
cm}^{-3}$, 
typically 1/3 of the energy is lost to neutrinos that escape the star,
while the other 2/3 of the energy is trapped inside the star.
Hence the luminosity from the DM heating is
\begin{equation}
\label{DMheating}
L_{DM} \sim {2 \over 3} \int \hat Q_{DM} dV 
\end{equation}
where $dV$ is the volume element.                                 

The properties of the collapsing protostellar clouds have been given
by 3D simulations (Abel et al. 2002; Gao et al. 2007).  At the time
when the density reaches $n = 10^{13}$cm$^{-3}$, the critical value
for 100 GeV particles, Paper I found a proto-DS in equilibrium with a
radius of 17 AU and a mass of $0.6 M_\odot$, giving a DM luminosity of
$\sim 140$ L$_\odot$. As more mass accretes onto the DS, the
protostellar luminosity begins to exceed the DM heating, so that the
protostar is no longer in thermal equilibrium.  Thus it must contract
which increases the DM density until the DM heating as given in
equation (\ref{DMheating}) matches its radiated luminosity.
%(hydrodynamic evolution in preparation).       

In this calculation, we assume that such a situation can be reached,
and we then build up the Dark Star from a few solar masses up to 1000
$M_\odot$, finding its structure as a polytrope in hydrostatic and
thermal equilibrium at each step in mass.  As we build up the star
more DM is pulled into the star via adiabatic contraction and
subsequently annihilates; we find that the annihilation fuel contained
in the star can thereby last $\sim 10^6$yr.  While the results of this
paper were being written, a paper appeared by Iocco et al. (2008)
which included DM heating in Pop III pre-main-sequence evolution of a
set of stars of {\it fixed mass}, finding that the quasi-hydrostatic
contraction is halted for times of $2 \times 10^3$ ($2 \times 10^4$)
yr for stars of mass 600 (9) M$_\odot$, at radii $\approx$ a few AU.

During the evolution of a DS, additional WIMPs could be captured via
scattering off of nuclei.  The cross section for scattering
($\sigma_s$) is very uncertain. For $\sigma_s <10^{-39}$ cm$^2$ we
find that a DM particle undergoes less than one scattering event in 1
Myr in the evolutionary stage considered in this paper. The
experimental bounds for 100 GeV particles from DM searches are
$\sigma_s \lesssim 2 \times 10^{-43}$ cm$^2$ for the spin-independent
case (Gaitskill et al. 2008) and $\sigma_s \lesssim 3.5 \times
10^{-39}$ cm$^2$ for the spin-dependent case (Savage et al. 2004).
Hence we assume negligible scattering here. However at later stages of
the evolution, once the DM density becomes too low to support the star
via heating, the DS contracts until nuclear burning sets in. At these
higher densities scattering at the experimentally allowed limit would
become important. DM passing through the star could be captured and
again drive DM heating.  These effects have been considered for
main-sequence and pre-main-sequence DS (Freese et al. 2008; Iocco
2008; Iocco et al. 2008), who find that the DM heating could dominate
nuclear fusion as long as the background DM density (from which the capture
takes place) remains high enough. Future work will further consider scattering in the
DS.

We also cite previous work on DM annihilation in today's stars 
(less powerful than in the first stars): Krauss et al (1985);
Bouquet \& Salati (1989); Salati \& Silk (1989); Moskalenko \& Wai (2007);
Scott et al. (2007); Bertone \& Fairbairn (2007).

\section{Equilibrium Structure}

We make the assumption that the dark
stars (DS) can be described as polytropes in hydrostatic equilibrium
\begin{equation}
\label{eq:polytrope}
P = K \rho^{1 + 1/n} .
\end{equation}
where $P$ is the pressure, $\rho$ is the density, and  the constant $K$ is
determined once the total mass and radius are specified (Chandrasekhar
1939).  Pre-main-sequence stellar models are adequately described by polytropes
in the range $n=1.5$ (fully convective) to $n=3$ (fully radiative). 
For a given stellar mass, we iterate the radius of the  model    
to find the point of thermal equilibrium, that is, 
the total DM  heating matches the radiated luminosity.
We then add  1 M$_\odot$, calculate a new equilibrium, and continue up to 
 1000  M$_\odot$.  In the standard scenario
of formation of the first stars, it was found that at $n \sim 10^4$
cm$^{-3}$, the mass of the protostellar cloud exceeds the Jeans
mass (e. g. Bromm, Coppi, \& Larson 2002). This amount of baryonic material, 
$\sim 1000 $ M$_\odot$, could fall  down
onto the DS and in the process bring in more DM with it. 
%In fact, even more matter could rain down, since the
%initial halo contains $\sim 10^5 M_\odot$
%in baryonic matter and $\sim 10^6 M_\odot$ in DM, but 
%we stop the sequence at 1000 M$_\odot$.  

\subsection{DM Densities}
The DM densities in the protostar are derived as described in Paper I.
We take a $10^6$ M$_\odot$ halo composed of 85\% DM and 15\% baryons.
We take an initial Navarro, Frenk, \& White profile (1996; NFW) with a
concentration parameter $c=2$ at $z=20$ in a standard $\Lambda$CDM
universe.  We follow the DM response to the changing baryonic
gravitational potential as the protostellar gas condenses.  As the
baryons come to dominate the potential well in the core, they pull the
DM particles inward. We use the simple adiabatic contraction method of
Blumenthal et al. (1986), Barnes \& White (1984), and Ryden \& Gunn
(1987) (hereafter Blumenthal method) to estimate the resultant DM
density profile.  The method has the limitation that all halo
particles are taken to be on circular orbits.  Recently (Freese et
al. 2008b), we did an exact calculation using an algorithm originally
developed by Young (1980) which takes into account radial motions as
well.  The results for the DM density agree with those from the
Blumenthal method to within a factor of 2.  This factor of 2 may be
compensated by the fact that recent simulations by Via Lactea II
(Diemand et al. 2008) find initial DM density profiles that are
steeper in the inner core ($\rho_\chi \propto r^{-1.2}$ rather than
$\rho_\chi \propto 1/r$).  Hence, the Blumenthal method should give
reasonable results.  The DM density profile in the DS is calculated at
each iteration of the stellar structure, so that the DM luminosity can
be determined.

\subsection{Basic Equations}

The basic equation is that of hydrostatic equilibrium 
\begin{equation}
{dP \over dr} = - \rho {GM_r \over r^2}
\end{equation} 
where ${dM_r \over dr} = 4 \pi r^2 \rho(r)$, $\rho(r)$ is the
total density (gas plus DM) at radius $r$, and $M_r$ is the enclosed mass
within radius $r$.
The temperature of the gas  $(T(r)$) is determined from the  equation of state 
of a  mixture of ideal gas and radiation:
\begin{equation}
\label{eq:eqnofstate}
P(r) = {{\rho k_B T(r)}\over {m_u \bar m}} + {1 \over 3} aT(r)^4 
= P_g + P_{rad}
\end{equation}
where $k_B$ is Boltzmann's constant, 
$m_u$ is the atomic mass unit,  and the mean atomic weight $\bar m
 = (2X + 3/4 Y)^{-1} =0.588$.  We take the H mass fraction $X=0.76$
 and the He     mass fraction $Y=0.24$.  In the resulting models   
 $T\gg10,\!000$ K except near the very surface, 
so  the H and He    are ionized and the H$_2$ is dissociated.
We will find the radiation pressure to be important once the DS
becomes heavier than $\sim 100 M_\odot$.
We also require the DS to be in thermal equilibrium,
\begin{equation}
L_* = 4 \pi \sigma_B R_S^2 T_{\rm eff}^4 = L_{DM} 
\label{eq:stellarlum}
\end{equation}
where $T_{\rm eff}$ is the effective surface temperature of the star at
its photospheric radius $R_S$.  Note that the observable properties
are determined by this temperature.
The location of the photosphere may be determined roughly by the
requirement that the optical depth outside of $R_S$ is $\tau \sim 1$,
which is equivalent to using the photospheric boundary condition
$\kappa P= {2 \over 3} g$ where $g$ is the surface gravity.  We use a
zero-metallicity Rosseland mean opacity ($\kappa$) table from OPAL
(Iglesias \& Rogers 1996), supplemented at the lowest temperatures by
opacities from Lenzuni et al. (1991).

For each DS mass, we find the equilibrium star in the following way.
We guess  a value for the outer radius  $R_S$ 
which, along with the mass and the polytropic assumption,
can then be used to determine the baryon density distribution. Then
the Blumenthal method      determines the DM density.  One can then
use equations (\ref{eq:Q}) and  (\ref{DMheating}) to find the amount of heating in each shell.
Our stellar code integrates outwards from the center of the DS, takes
a few hundred radial steps, and stops once it satisfies the photospheric
boundary condition. The temperature there is set to $T_{\rm eff}$.
Now, one compares $L_\ast$ with $L_{\rm DM}$.
If $L_{DM}<L_*$, then the next guess for $R_S$ must be smaller in order to
increa
boundary condition. The temperature there is set to $T_{\rm eff}$. 
Now, one compares $L_\ast$ with $L_{\rm DM}$.
If $L_{DM}<L_*$, then the next guess for $R_S$ must be smaller in order to
increase $L_{DM}$ and at the same time decrease $L_\ast$. 
 Conversely, if $L_{DM} > L_\ast$, then the star must expand
in order to reduce the DM heating.  We iterate to a convergence in the $L$'s
to 1 part in $10^4$.

\subsection{Building up the Mass}

Then we allow surrounding matter from the original baryonic core
 to accrete onto the DS
at  $2 \times 10^{-3}$
M$_\odot/{\rm yr}$, roughly $M_{\rm core}/t_{\rm ff}$, where $t_{\rm ff}$
is the free-fall time of the core.
 The initial  DS mass is 3 M$_\odot$ and the increment is 
1 M$_\odot$. We   
remove the amount of DM that has annihilated at each stage at each
radius.  We continue  stepping up in mass until we
reach 1000 M$_\odot$, 
the Jeans mass of the core      (Bromm \& Larson 2004). 

With the above accretion rate, it takes $5 \times 10^5$ yr to build up
 to 1000 M$_\odot$.  Hence the lifetime of the DS is at least this
 long.  By this time a significant fraction of the DM inside the DS
 has annihilated away.  It is not
 known whether or not the DM inside the DS can be repopulated from DM
 particles in the $10^6 M_\odot$ halo surrounding it; this question
 would require numerical resolution not currently available.

In the future, it would be interesting to study the accretion process
in more detail.  It is likely to proceed via the formation of a disk
with an accompanying accretion luminosity.  In the standard Pop III
star formation process of accretion onto a small $10^{-3}M_\odot$
nugget, the luminosity has an accretion-driven phase; here, on the
other hand the accretion luminosity of the much larger DS is
always negligible.  In any case our treatment of the structure of the
stellar interior is probably unchanged by the presence of the disk.
Previously McKee \& Tan (2007) have studied the role of angular
momentum in Pop III stars in the absence of DM.  One should reconsider
angular momentum in the case of DS as well.

\section{Results}

Table 1 and Figure 1 illustrate our results for standard parameters 
for $M_*=(10 - 1000) M_\odot$ and for n=1.5.  In the Table we present the sequence
of central temperature $T_c$,  photospheric
radius $R_S$, central gas density $\rho_c$, central DM density
$\rho_{\chi,c}$, stellar luminosity (equal to DM heating luminosity)
$L_*$, surface temperature $T_{\rm eff}$, total DM mass inside the star
$M_{DM}$, and time evolved since DM heating dominates inside the star.
At 1000 $M_\odot$  $\rho_c$
is far lower than for any metal-free Zero Age Main
Sequence star.
Figure 1 plots the baryon and DM density profiles.
 The DM density is many of orders of
magnitude lower than the baryon density throughout the evolution
and yet the DM annihilation powers the star.  As time goes on, one
can see that the DM is depleted in the interior regions of the star,
due to annihilation, and the density becomes very nearly constant.
The plot of $L_{\rm DM}(r)$, the dark matter luminosity integrated out to
radius $r$, shows that the heating is spread out over much of 
the volume of the DS; thus it is not particularly sensitive to changes
in the details of the adiabatic contraction model. 
By the time the DS reaches $1000$ M$_\odot$, the
amount of DM in the star is only 1/3 $M_\odot$, and 1/3 of the DM in
the DS has annihilated away. We also find that at each evolutionary state,
 $L_\ast$ is typically
an order of magnitude less than the Eddington luminosity for a star
of that mass and radius.

We have found that the DS is fully convective for stellar masses below
$100 M_\odot$; makes the transition from convective to radiative in
the $M_*=(100-200) M_\odot$ mass range (with a radiative zone growing
outward from the center); and then becomes (almost) fully radiative
(but for a small convective region at the surface) for $M_* > 200
M_\odot$.  Hence the initial convective period is best described by an
$n$=3/2 polytrope while the later radiative period is best described
by an $n$=3 polytrope.

The results above are for an $n$=3/2 polytrope all the way up to the
final mass.  For an $n$=3 polytrope (more appropriate at the higher
masses), calculated up to 600 $M_\odot$, the results are
qualitatively the same.  For $M_*=600 M_\odot$, the $n$=3 case gives
$T_{\rm eff}=9100K$, $R_S= 6.0 \times 10^{13}$cm, $L_*=4.6 \times 10^6
L_\odot$, and $T_c =2.2 \times 10^6K$; while the $n$=1.5 case gives
$T_{\rm eff}=6370K$, $R_S=1.0 \times 10^{14}$cm, $L_*=3.04 \times 10^6
L_\odot$, and $T_c = 6.88 \times 10^5K$.  Thus the results for the
$n=1.5$ polytrope give the basic picture.  Note that DS have much
lower $T_{\rm eff}$ than their standard metal free (Pop. III)
main-sequence counterparts in the absence of DM, which radiate at
$T_{\rm eff}>$30,000K. This difference gives a markedly different
observable signature for the DS than for the standard Pop III stars.

\section{Conclusions}

We have followed the growth of equilibrium Dark Stars,
powered by DM annihilation, up to 1000 M$_\odot$. The objects have
sizes of a few AU and central $T_c\approx 10^5-10^6$ K.  Sufficient DM
is brought into the star by contraction from the DM halo to result in
a DS which lives at least 0.5 Myr  (the lifetime could be significantly
longer if DM capture becomes important
at the later stages, as long as the background DM density is high
enough for capture to take place). Because of the relatively low $T_{\rm
eff}$ (4000--10,000 K), feedback mechanisms for shutting off accretion
of baryons, such as the formation of HII regions or the dissociation
of infalling H$_2$ by Lyman-Werner photons, are not effective.  The
implication is that main-sequence stars of Pop. III are very massive.
This conclusion depends on uncertain parameters such as the DM
particle mass, the accretion rate, and scattering, effects that will
be studied in future work.
                
Although DS shine with a few $10^6 L_\odot$ they would be very
difficult to observe at $z \sim 10-50$. One can speculate that
pristine regions containing only H and He might still exist to lower
redshifts; then DS forming in these regions might be easier to detect.
One may hope that the ones that form most recently are detectable by
JWST or TMT and differentiable from the standard metal-free Pop. III
objects.  DS are also predicted to have atomic hydrogen lines
originating in the warmer photospheres, and H$_2$ lines arising from
the infalling material, which is still relatively cool.

It has been argued that Pop III.1 stars (the very first metal-free
 stars) may constitute at most $\sim 10\%$ of metal poor stars on
 observational grounds.  Heger \& Woosley (2002; HW) showed that for
 $140 M_\odot < M < 260 M_\odot$, pair instability (SN) lead to
 odd-even effects in the nuclei produced that are strongly constrained
 by observations.  Thus if Pop III.1 stars are really in this mass
 range one would have to constrain their abundance. For $M > 260
 M_\odot$, HW find that no SN occurs, and the end result of stellar
 evolution is collapse of the entire star into a black hole. We
 expect, based on extension of the $n=3$ calculation, that our DS runs
 out of DM at about 700--900 M$_\odot$ (for $m_\chi = 100$ GeV).  Then
 it must contract to the main sequence, where nuclear burning sets in,
 and further evolution would proceed as in HW. Alternatively, the
 evolution could proceed as described by Ohkubo et al. (2006) who
 found that metal-free stars of 500 and 1000 $M_\odot$, taking into
 account two-dimensional effects, did blow up as SN, leaving about
 half their mass behind in a black hole.  In this case the SN might be
 observable signatures of DS, distinguishable since they arise from
 such high mass stars.  The end product in either case would be a
 plausible precursor of the otherwise unexplained $10^9 M_\odot$ black
 holes at $z=6$ (Yoshida et al., in prep).

We acknowledge support from: the DOE and MCTP via the Univ.\ of
Michigan (K.F.); NSF grant AST-0507117 and GAANN (D.S.); NSF grant
PHY-0456825 (P.G.). K.F. acknowledges the hospitality of the Physics
Dept. at the Univ. of Utah.  K.F. and D.S. are extremely grateful to
Chris McKee and Pierre Salati for their encouragement of this line of
research, and to A. Aguirre, L. Bildsten, R. Bouwens, J. Gardner,
N. Murray, J. Primack, M. Rieke, C. Savage, J. Sellwood, J. Tan, and
N. Yoshida for helpful discussions.

\clearpage

\begin{table*}
 \caption{ Properties and Evolution of Dark Stars for $m_\chi = 100$
GeV, 
$\dot M = 2 \times 10^{-3} M_\odot/{\rm yr}$, $\langle \sigma v \rangle = 3 \times 10^{-26}$ cm$^3$/s,
polytropic $n= 1.5$. 
%\vspace{0.1 in}
}
\begin{center}
{
\small
 \begin{tabular}{||l|l|l|l|l|l|l|l|l|c||}
 \hline\hline
$M_*$ & $T_c$ & $R_S $ & $\rho_c $ & 
$\rho_{\chi,c} $ & $L_* $ & $T_{eff}$ 
& $M_{DM}$ & t \\ 
$(M_\odot)$ & $(10^5 {\rm K})$ & $(10^{13} {\rm cm})$ & $ ({\rm gm/cm}^3)$ & 
$({\rm gm/cm}^3)$ & $(L_\odot)$ & $(10^3 {\rm K})$ 
& (gm)  & (yr)  \\ 
\hline
  12 
& $1.3$
& $4.2$
& $4.1 \times 10^{-7}$
& $1.1 \times 10^{-9}$
& $1.1 \times 10^5$
& $4.3$
& $2.8 \times 10^{31}$
& $6 \times 10^3$ \\
\hline
  50
& $2.7$
& $6.0$
& $6.2 \times 10^{-7}$
& $1.2 \times 10^{-9}$
& $4.2 \times 10^5$
& $5.0$
& $9.1 \times 10^{31}$
& $2.5 \times 10^4$ \\
\hline
  100
& $3.5$
& $7.1$
& $7.7 \times 10^{-7}$
& $1.1 \times 10^{-9}$
& $7.8 \times 10^5$
& $5.3$
& $1.6 \times 10^{32}$
& $5 \times1 0^4$ \\
\hline
  300 
& $5.3$
& $9.0$
& $1.2 \times 10^{-6}$
& $8.2 \times 10^{-10}$
& $1.9 \times 10^6$
& $6.0$
& $3.6 \times 10^{32}$
& $1.5 \times 10^5$ \\
\hline
  1000 
& $8.5$
& $10$
& $2.4 \times 10^{-6}$
& $4.5 \times 10^{-10}$
& $3.9 \times 10^6$
& $6.6$
& $7.3 \times 10^{32}$
& $5\times10^5$
\\
 \hline 
 \hline\hline
 \end{tabular} 
 \label{tab:ExpParam}
}
\end{center}
\end{table*}

\clearpage

\begin{figure}[t]
%\epsscale{0.7}
\plotone{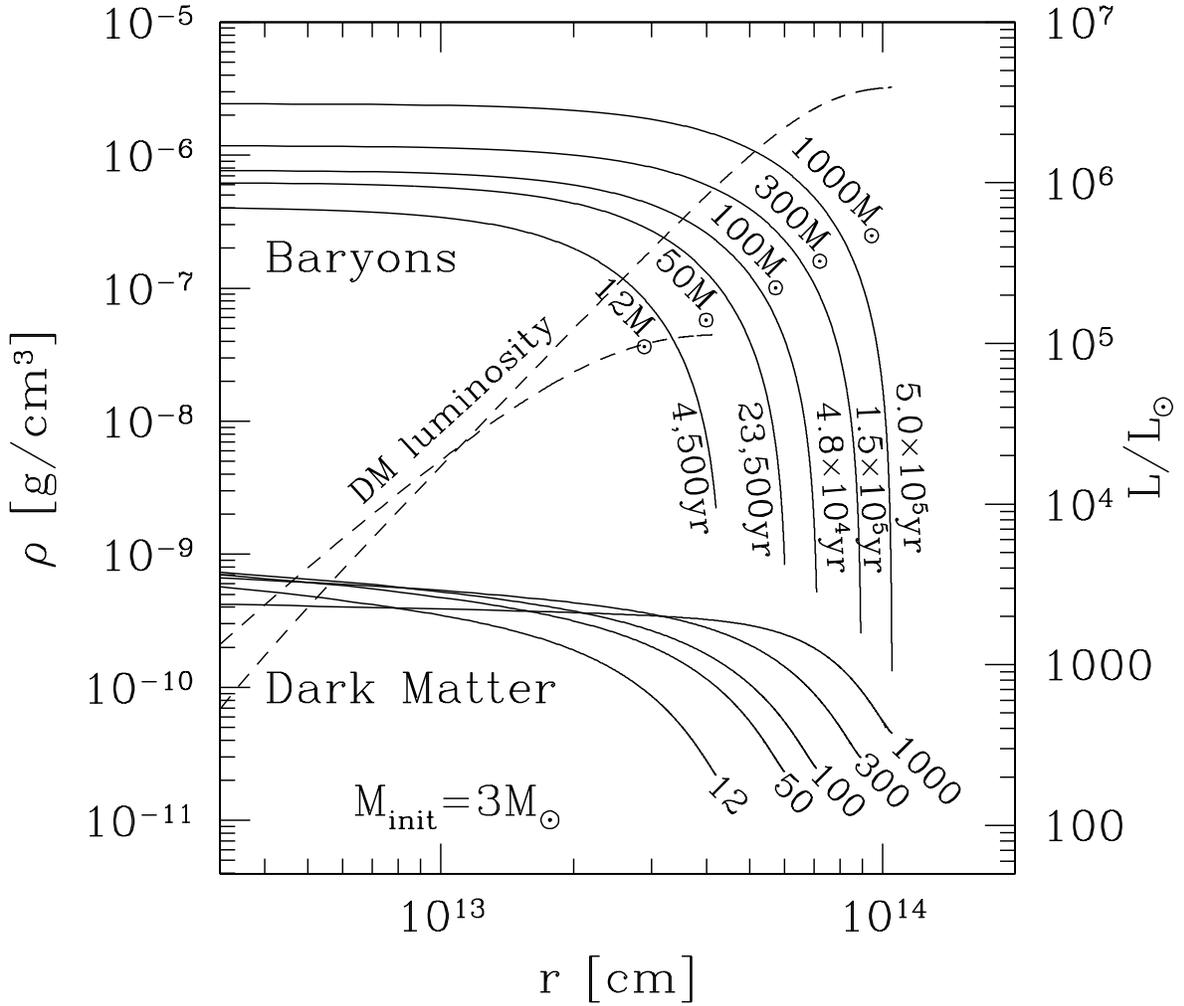}
%\centerline{\includegraphics[width=0.4\textwidth]{f1.eps}}
\caption{Evolution of a dark star (n=1.5) as mass is accreted onto the initial
protostellar core of 3 M$_\odot$.  The set of upper
(lower) solid curves correspond to the baryonic (DM) density profile (values given
on left axis) at different
masses and times. {\it Dashed lines: } luminosity $L_{\rm DM}$ integrated out to radius $r$
for the masses 12 and 1000 M$_\odot$, in solar units (values given on the right axis).
%After $ 5 \times 10^4$ ($5 \times 10^5$) years, the DS
%mass reaches $100 (1000) M_\odot$.  The DM profile has been
%consistently adjusted to account for DM annihilation.  As a result,
%the central DM density has decreased relative to an early time in its
%evolution.
%\vspace{0.1 in}
}
\end{figure}

\end{document}